\documentstyle[psfig]{caosp}
\begin{document}
\def\Hm{\langle H\rangle}
\def\Hs{(\Hsq+\Hzsq)}
\def\Ht{\Hs^{1/2}}
\def\Hsq{\langle H^2\rangle}
\def\Hzsq{\langle H_z^2\rangle}
\def\HgII{\hbox{{\rm Hg}~{$\scriptstyle {\rm II}$}}}
\def\PtII{\hbox{{\rm Pt}~{$\scriptstyle {\rm II}$}}}
\def\AuII{\hbox{{\rm Au}~{$\scriptstyle {\rm II}$}}}
\def\MgII{\hbox{{\rm Mg}~{$\scriptstyle {\rm II}$}}}
\def\fy{\hbox{$.\!\!^{\rm y}$}}
%
%
%%%%%%%%%%%%%%%%%%%%%%%%%%%%%%%%%%%%%%%%%%%%%%%%%%%%%%%%%%%%%%%%%%%%%%%%%%%%%%
%                       E D I T O R I A L   N O T E S                        %
% Next 3 lines are not obligatory, they will be inserted by the editors      %
%%%%%%%%%%%%%%%%%%%%%%%%%%%%%%%%%%%%%%%%%%%%%%%%%%%%%%%%%%%%%%%%%%%%%%%%%%%%%%
\pubyear{1993}
\volume{23}
\firstpage{1}
%%%%%%%%%%%%%%%%%%%%%%%%%%%%%%%%%%%%%%%%%%%%%%%%%%%%%%%%%%%%%%%%%%%%%%%%%%%%%%
\htitle{HgMn stars: new insights}
\hauthor{S. Hubrig {\it et al.}}
%%%%%%%%%%%%%%%%%%%%%%%%%%%%%%%%%%%%%%%%%%%%%%%%%%%%%%%%%%%%%%%%%%%%%%%%%%%%%%
%                       T I T L E                                            %
% You should  not use just capital letters in the title. Don`t end the       %
% title by "." (dot)                                                         %
%%%%%%%%%%%%%%%%%%%%%%%%%%%%%%%%%%%%%%%%%%%%%%%%%%%%%%%%%%%%%%%%%%%%%%%%%%%%%%
\title{HgMn stars: new insights}
%%%%%%%%%%%%%%%%%%%%%%%%%%%%%%%%%%%%%%%%%%%%%%%%%%%%%%%%%%%%%%%%%%%%%%%%%%%%%%
\author{S. Hubrig \inst{1}}
%%%%%%%%%%%%%%%%%%%%%%%%%%%%%%%%%%%%%%%%%%%%%%%%%%%%%%%%%%%%%%%%%%%%%%%%%%%%%%
\institute{University of Potsdam, Am Neuen Palais 10, D-14469 Potsdam, 
Germany}

%%%%%%%%%%%%%%%%%%%%%%%%%%%%%%%%%%%%%%%%%%%%%%%%%%%%%%%%%%%%%%%%%%%%%%%%%%%%%%
%\date{\today}
\maketitle
%%%%%%%%%%%%%%%%%%%%%%%%%%%%%%%%%%%%%%%%%%%%%%%%%%%%%%%%%%%%%%%%%%%%%%%%%%%%%%
\begin{abstract}
Recent results obtained by various authors on the properties
of HgMn stars are reviewed. Substantial progress has been achieved in the
study of abundances and isotopic anomalies. The results about the magnetic 
fields
and membership in multiple systems suggest further directions of investigations
to be followed in view of answering the question of the development of 
abundance peculiarities in HgMn stars.
\end{abstract}
%%%%%%%%%%%%%%%%%%%%%%%%%%%%%%%%%%%%%%%%%%%%%%%%%%%%%%%%%%%%%%%%%%%%%%%%%%%%%%
\section{Introduction}
%%%%%%%%%%%%%%%%%%%%%%%%%%%%%%%%%%%%%%%%%%%%%%%%%%%%%%%%%%%%%%q%%%%%%%%%%%%%%%%
\label{intr}
Recent years have seen a renewed interest in HgMn stars. 
The main objective of studies of HgMn stars is
to provide new, better observational data for the theorists investigating the 
mechanisms responsible for abundance anomalies in these stars.  
Much spectroscopic work was devoted to elemental abundance analyses.
About 30 papers on abundances in individual stars and in samples of stars 
appeared 
in the last three years. Good studies of the correlations between
elemental abundances and
fundamental parameters are crucial for the
understanding of the physical processes taking place in HgMn stars.
On the other hand, in the consideration of the physical mechanisms 
contributing to the development of the wide range of 
abundance and isotopic anomalies, it is important to know the r\^ole that 
magnetic fields play.
 
Here I wish to discuss some of the observations that deal with
the problems of
anomalous abundances of the heavy elements Hg and Pt and with the 
question of the presence of surface magnetic fields. Finally, I shall 
discuss the results of recent statistics of multiple systems among HgMn stars. 
%%%%%%%%%%%%%%%%%%%%%%%%%%%%%%%%%%%%%%%%%%%%%%%%%%%%%%%%%%%%%%%%%%%%%%%%%%%%%%
\section{Abundance studies}
Significant progress in abundance determinations has been achieved
in the last years by obtaining observations
of increased spectral resolution.
The best accurate quantitative studies in the complex ultraviolet region were
done with the Goddard High Resolution Spectrograph (GHRS) on the Hubble 
Space Telescope
(HST) (Leckrone et al. 1998). The data obtained from space for a small 
number of HgMn stars has led to the identification of exotic species 
such as Ge, As, Ru, Rh, Pd, 
Cd, Os, Tl, Pb, Bi, and to the quantitative
determination of their abundances.
Significant
progress in the study of chemical abundances and isotopic anomalies
in HgMn stars was made possible by the availability of new atomic data,
including laboratory measurements and computations
of wavelengths, oscillator strengths, isotope shifts, and hyperfine structure
for a variety of elements in different ionization stages. 
Close collaboration between members of
the HST team and atomic physicists allowed the nature of chemical 
anomalies in the UV region of HgMn stars to be unveiled.

An important result achieved through the studies from space is the
finding of discrepancies in abundances derived from different 
ionization  states of the
elements Si, S, Co, As, Y, Zr, Pt, Au and Hg.
Such abundance discrepancies among ionization states
probably result from a vertically stratified distribution of the given 
element within the observable outer layers of the star. However, it cannot be
completely excluded that these discrepancies
reflect errors in atomic data or are caused by departures from the LTE 
ionization equilibrium.
The heavy-element peak in the best studied star, {$\chi$~Lup} (= HD 141556),
rises sharply to Pt, Au, Hg and Tl, and then falls 
steeply to Pb and Bi (Leckrone
et al. 1998). 
The isotopic anomalies were studied for the heavy 
elements Hg, Pt, Tl and Pb. Exciting results
were found from GHRS observations of the isotopic shifts in the platinum 
and mercury lines in the two sharp-lined stars {$\chi$~Lup} and HR 7775.
For the star {$\chi$~Lup}, platinum consists of a 50--50 mixture 
of isotopes $^{196}\!$Pt and $^{198}\!$Pt (Kalus et al. 1998). 
This result is discrepant with what had been found long ago from 
a study in the optical region
by Dworetsky \&\ Vaughan (1973). They had reported that Pt was
present as essentially pure $^{198}\!$Pt, the heaviest isotope.
Without laboratory data for isotope shifts in Pt, no spectrum synthesis could 
be done at that time. 
Recent observations of {$\chi$~Lup} in the visual at very high resolution,
R = $\lambda/\Delta\lambda\ga100\,000$ (Jomaron et al., Hubrig et al.,
these
proceedings), combined with new data for the wavelengths of isotopic and
hyperfine components of \PtII\ (Engleman 1989), confirm the concentration 
of Pt in its heaviest isotope.

As for {$\chi$~Lup}, the identified platinum isotope mixture 
at optical wavelengths for the HgMn star HR 7775 appears to be different from 
that obtained from 
ultraviolet transitions (Wahlgren \&\ Dolk, these proceedings; Kalus et al.
1998). Wahlgren \&\ Dolk reported that in this star Hg isotopic 
shifts vary with the ionization stage.

In agreement with previous works on Hg and Pt
(White et al. 1976; Dworetsky \&\ Vaughan 1973; Smith 1997),
new determinations of isotopic anomalies in the sample of sharp-lined HgMn 
stars in the
optical region show that
in the cooler stars both Hg and Pt are concentrated in their heaviest 
isotopes (Jomaron et al., Hubrig et al., these proceedings). However, 
detailed
abundance structure of the isotopes does not follow the fractionation model 
introduced by White et al. (1976) and varies from star to star.

Presently, the principal difficulty involved in 
interpreting the behavior of 
the isotopes in HgMn stars is that no combination of 
gross stellar physical parameters is sufficient to characterize the observed 
variations of isotopic composition. 

In recent years a number of papers on quantitative analyses of 
individual HgMn 
stars and
of samples of such stars from spectra of moderate resolution have 
appeared.  
One survey, based mostly on archival IUE spectra, has been devoted to 
a study of 26 HgMn stars and 14 normal late B stars, to investigate the 
group characteristics (Smith \&\ Dworetsky 1993; Smith 1993, 1994,
1995, 1997). 
The abundances of several elements were shown to be
correlated with stellar effective temperature, which suggests that diffusion
is operating to modify the atmospheres of HgMn stars.
On the other hand, it is not clear how the abundances in HgMn stars are 
affected by stellar rotation. If we expect that 
diffusion is operating to modify the atmosphere of HgMn stars, then
the extent of abundance anomalies should progressively decrease 
with increasing  rotational velocity. This has not been demonstrated yet.
Additional abundance analyses of HgMn stars with different 
$v\,\sin i$ values may 
help to clarify whether there is any correlation of abundance with rotational 
velocity within the class of HgMn stars. 
The recent studies of Smith \&\ Dworetsky and Smith  
once again showed that the HgMn phenomenon is 
heterogeneous, i.e. within the overall class there are subgroups with 
star-to-star diversity in abundances (see e.g., Sargent \&\ Searle 
1967).

In the past, diffusion has proven a successful mechanism
to explain the overall trends of
surface compositions of chemically peculiar stars.
A challenge to any theory now is to identify a mechanism that accounts for 
star-to-star variations in chemical composition and isotopic
abundance.

\section{Magnetic fields}
Since no combination of stellar parameters has been found to be sufficient
to characterize the variety of the surface composition on HgMn stars, 
some unrecognized factor should be involved 
among the necessary and sufficient 
conditions for occurrence of the HgMn phenomenon.
Until recently, the issue of the presence of a magnetic field
in HgMn stars remained open, and the r\^ole that
magnetic fields possibly play 
in the development of anomalies in these stars had
never been critically tested. This  
is a fundamental question whose answer is essential for the understanding
of the physical processes taking place in HgMn stars and, more generally, 
during the formation and evolution of B stars.  

Previous searches for magnetic fields in HgMn stars had shown that these stars,
unlike classical Ap stars, do not have large-scale organized fields detectable
through spectropolarimetry. It has never been ruled out, though, that they 
might have tangled magnetic fields of the order of a few thousand gauss with no
net longitudinal component.
  
The reinvestigation of the magnetic fields at the present time appears to be 
promising for several reasons. 
 
First, Mathys \&\ Lanz (1990) suggested that some Am stars
may have magnetic fields with a structure more complex than that of the
classical Ap stars. The detection of a magnetic field of about 2~kG in the
hot Am star {$o$} Peg (= HD 214994) through the Stenflo-Lindegren
multi-line technique, using the differential broadening of lines having 
different Zeeman sensitivities raised the question whether the hotter
counterparts of Am stars, the HgMn stars, also have magnetic fields with a 
complex structure not detectable through spectropolarimetry.

Second, the recent development of the moment technique by Mathys (1988, 1993, 
1995) has
significantly 
simplified the problem of determination of fields with no net longitudinal 
component. 
Successful application of this method to the determination of parameters
related to the magnetic field have been presented in a number of papers
(Mathys 1993; Mathys 1995; Mathys \&\ Hubrig 1997). 

Third, the diagnostic potential of 
circularly polarized spectra recorded with the Zeeman analyzer of the
Cassegrain Echelle Spectrograph (CASPEC) at the ESO 3.6 m telescope has
considerably improved. 
From 1993, the instrumental setup has been modified (see for more 
details Mathys \&\ Hubrig 1997). The configuration used before allowed to 
achieve a spectral resolution of R = $\lambda/\Delta
\lambda=18\,000$.
With the new configuration spectra can be
recorded at a resolving power R = $\lambda/\Delta
\lambda=39\,000$  over the range 5600-6800~\AA. The
moment technique could be applied to analyze the shapes of spectral
lines in high-resolution low-noise spectra.

Our first results of the measurements of magnetic fields in the two SB2 stars 
74~Aqr (= HD 216494) and {$\chi$~Lup} with the Zeeman analyzer of CASPEC 
were published two years ago (Mathys \&\ Hubrig 1995). We applied the
moment technique to look for possible differential broadening of spectral 
lines having different magnetic sensitivities. In this approach, the second 
order moments of the 
profiles of a sample of lines of a single ion (namely, Fe II) were measured and 
their dependence on the atomic parameters characterizing the magnetic line 
broadening was studied. Our study revealed that the HgMn 
primary of the system 74~Aqr has a quadratic field of $(3.6\pm0.8)$~kG. 
We have also been able to detect a longitudinal field of a few hundred gauss
in the secondary component of {$\chi$~Lup} at the 
$4.6\,\sigma$ level. No quadratic field was detected for {$\chi$~Lup}. 
However, the standard error of the determination was rather large, 1.25~kG.
In follow-up observations of the 
HgMn stars HD 27376, HD 78316, HD 174933, 
HD 191110 and  {$\chi$~Lup} scheduled two years later, we could measure
longitudinal fields of the order of few hundred
gauss at levels above $3\,\sigma$ (Hubrig \&\ Mathys 1998). No quadratic
field detection could be achieved at a significant level.

Nevertheless, these results do not rule out that the 
considered stars might 
have tangled fields.
The diagnosis of the quadratic field is difficult and it depends much more
critically on the size of the set of lines that can be employed than the 
diagnosis of the longitudinal field.
A straightforward continuation in the study of magnetic fields in HgMn stars
would be to reduce the measurements uncertainties. This can be achieved by
increasing the size of the sample of lines used to diagnose the magnetic field,
that is, observing a wider spectral range, and by achieving better spectral
resolution.
In 1996 we could take advantage of the unique performance 
of the high-resolution echelle mode of the ESO Multi-Mode
Instrument (EMMI) at the New Technology Telescope (NTT) on La Silla. Using
this instrument, both the wavelength coverage and the resolving power are 
double of those previously achieved with CASPEC. Out of three HgMn stars 
observed
with this configuration,
one, HD 78316, had a quadratic field of 2.7~kG at the $6.3\,\sigma$ level. 

The fact that only weak fields are observable in circular polarization 
while quadratic magnetic fields of kG order are probably present
most likely means 
that the field
has a fairly complex structure, such that in disk-averaged observations, the
contribution of various regions of the stellar surface to the net polarimetric
signal mostly cancels out. This does not necessarily imply a very 
discontinuous field structure like in the sun. Some types of
large-scale organization, such as high-order multipoles or toroidal
fields (which appear as predominantly transversal), may lead to the
observed cancellation. The latter may furthermore 
be very easily built up in binary systems. 

The group of HgMn stars is rich in moderately close binaries, with 
a frequency of more than 60\%\ (Hubrig \&\ Mathys 1995). 
It was shown by Zahn (1977) that the tidal forcing of an early-type star in 
a binary system excites gravity
waves at the boundary between the convective core and the radiative envelope.
He argued that the torque associated with the excitation of 
gravity waves is much greater than that associated with turbulent viscosity
in the convective core.
Zahn (1984) and Goldreich \&\ Nicholson (1989) examined 
in more detail the manner in which tidal spin-down proceeds in early-type 
stars. They conclude that tidal despinning to synchronous rotation proceeds
from the outside toward the inside of the star. 
The tidal torque per unit mass varies with depth and
latitude in a star, and therefore it tends to induce differential rotation.
 
We can assume that the rotation observed at the surface of the HgMn stars in 
binary systems could well be a poor indicator of the total angular momentum 
I{$\Omega$} stored in those stars and that the angular velocity is decreasing
outward.  
Because of the ubiquity of magnetic fields in the universe,
we can presume that a
weak poloidal field is present in these stars. Then, the toroidal field is 
produced from the poloidal field by the non-uniform rotation in a 
characteristic time of only years.
The toroidal fields should be able to diffuse slowly outward, 
or they might be 
raised to the stellar surface by eruption or meridional flows.
 
Some binary systems with an HgMn primary are known to have components which
definitely rotate subsynchronously (Guthrie 1986) and it is not clear yet 
how subsynchronous
rotation could be achieved. The detection of magnetic fields in a few HgMn
stars makes plausible the idea that, like in classical Ap stars,
a stellar magnetic field coupling the star with
its surrounding can transfer its angular momentum outside the star,
and so slow down rotation.
 
To understand better the nature of HgMn stars, further searches for magnetic 
fields will be worthwhile. To determine the 
structure of those fields, the possible variations of 
the detected fields should
be studied. 

\begin{figure}[t]
\psfig{file=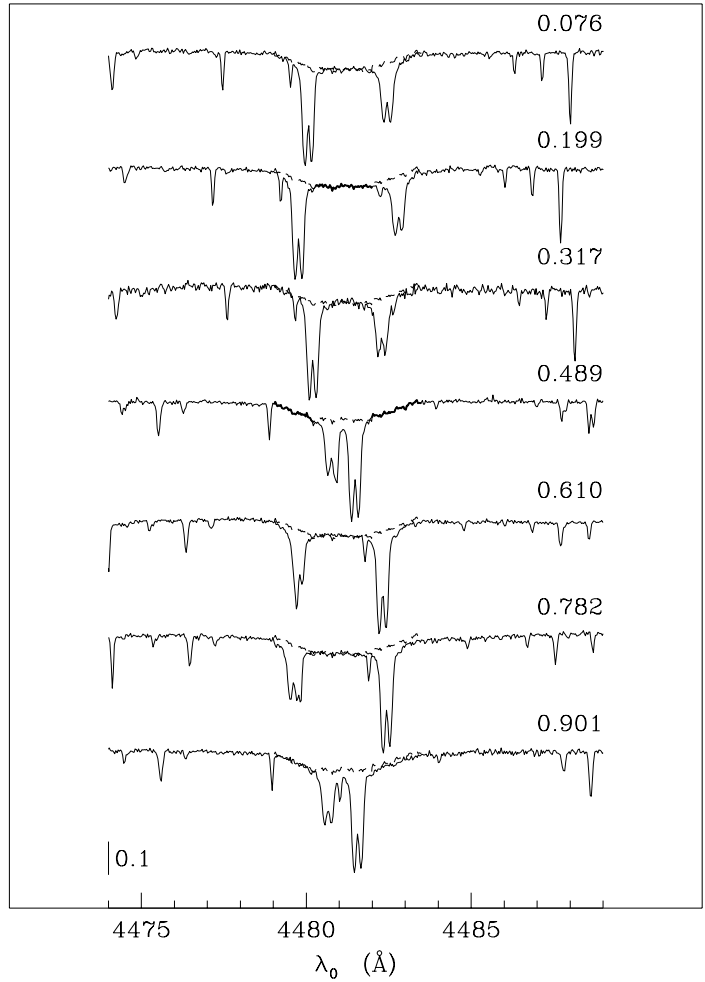,width=11.0cm}
\caption{The triple system 74~Aqr at various orbital phases}
\end{figure}

\section{ Multiplicity}
The last aspect I would like to address is the membership of HgMn stars
in multiple systems (i.e., systems with more than two components).
The HgMn stars have an unusually high proportion of multiple systems (Hubrig
\&\ Mathys 1995).
In the presently available catalogue of multiple stars by Tokovinin (1997), 
which compiles data on 612 stellar systems of different spectral types,
I have found four additional systems containing HgMn stars:
HD~32964, HD~36881, HD~58661 and HD~172044. 
The HgMn stars are restricted to the spectral range B6 to B9.
It is intriguing that if 
the relative frequency of HgMn stars in multiple systems 
is studied, every third system 
with the primary in this spectral range involves an HgMn star.

These observational results clearly show that the study of 
multiple systems with an HgMn component is of prime interest
for the star formation 
theories. To find out which r\^ole membership of HgMn stars in multiple systems
plays for the development of their chemical peculiarities, it would be  
important to compare the ranges of periods, luminosity ratios, 
and orbital eccentricities, as well as hierarchy of multiples with the same 
characteristics as normal late B systems.
Especially enlightening will be the determination of chemical 
composition of
the companions in such systems, as well as the search for possible relations 
between the magnetic fields and the abundance patterns of the various 
components.

On the other hand, it is still an intricate task to determine 
the fundamental parameters and chemical composition of the components in
binary and multiple systems. One triple system, 74~Aqr, has been presented 
by Hubrig \&\ 
Mathys (1994).
This system was observed at a resolving power 
R = $\lambda/\Delta\lambda=100\,000$ in 
the spectral region around the line
\MgII~$\lambda\,4481$ at various
orbital phases (Fig.~1). 
They are identified in the figure close to 
the
corresponding tracing. Thin lines indicate the contributions of the long-known
components of the binary, and broken lines those of the third component.
To define the central part of the latter, we have adopted the region 
comprised between the narrow components at phase 0.199, while its wings have
been taken from the regions on each side of the narrow components at
phase 0.489 (this procedure is visualized by thick line).  
The broad feature underlying the strong, sharp Mg II lines 
of both components appears constant along two orbital periods and seems to be
due to a third component. This third component can likely be identified as 
the companion observed through speckle interferometry, which has an orbital
period of 19\fy25 (Tokovinin 1993). Our observations suggest that its spectral
type is similar to or slightly earlier than that of the HgMn primary. However,
we have no clue about its possible peculiarity.
\section{Conclusions}
There has been considerable progress in recent years in our understanding of 
HgMn stars. However, many gaps still remain in our knowledge of how HgMn 
stars form and evolve. 
The mechanisms responsible for abundance anomalies have not been
definitely identified yet. Some systematic trends in the abundance data 
qualitatively support the mechanism of radiatively-driven diffusion. 
However, it is difficult to determine the mechanisms responsible for 
abundance anomalies in the absence of accurate observational information about
magnetic fields. It is, therefore, essential to settle the issue of the 
presence of magnetic fields in HgMn stars.

A potentially 
fruitful area for future research will be the abundance analyses of the
components of binary and multiple systems. We have still only relatively rough 
ideas about the properties of the companions in the systems. On the other
hand, such studies are difficult to carry out because 
most lines of many of the 
components of a double, triple 
or quadruple system appear quite weak (2-3\%\ deep) in the spectrum, 
as a result of the dilution by superposition of their continua. 
To my knowledge, such an
abundance analysis has until now been done only for one 
multiple system, HD~11905 (Zakharova 1997). 
%%%%%%%%%%%%%%%%%%%%%%%%%%%%%%%%%%%%%%%%%%%%%%%%%%%%%%%%%%%%%%%%%%%%%%%%%%%%%%
%\begin{table}[t]
%\small
%\begin{center}
%%%%%%%%%%%%%%%%%%%%%%%%%%%%%%%%%%%%%%%%%%%%%%%%%%%%%%%%%%%%%%%%%%%%%%%%%%%%%%


\begin{thebibliography}{}
%%%%%%%%%%%%%%%%%%%%%%%%%%%%%%%%%%%%%%%%%%%%%%%%%%%%%%%%%%%%%%%%%%%%%%%%%%%%%%
\article{Dworetsky, M.M., Vaughan, A.H.}{1973}{\apj}{181}{811}
\article{Engleman, R.J.}{1989}{\apj}{340}{1140}
\article{Goldreich, P., Nicholson, P.D.}{1989}{\apj}{342}{1079}
\article{Guthrie, B.N.G.}{1986}{\mnras}{220}{559}
\bibitem{}Hubrig, S., Mathys, G.: 1994, Poster paper presented
  at the JD No.~12 of the 22nd IAU General Assembly
\bibitem{}Hubrig, S., Mathys, G..: 1998, (in preparation)
\bibitem{}Hubrig, S., Mathys, G.: 1995, Comments Astrophys. {\bf 18}, 167
\bibitem{} Kalus, G., Johansson, S., Wahlgren, G.M., Leckrone, D.S.,
Thorne, A.P., Brandt, J.C. {\it Astrophys. J.}, in press 
\inproceedings{Leckrone, D.S., Johansson, S.G., Wahlgren, G.M., Proffitt, C.R.,
Brage, T.}{1998}{GHRS Science Symposium}{J.~Brandt, C.C.~Petersen \and
T.~Ake}{ASP Conf. Series}{San Francisco}{in press}
\article{Mathys, G.}{1988}{\aaa}{189}{179}
\inproceedings{Mathys, G.}
{1993}{Peculiar Versus Normal Phenomena in 
A-Type and Related Stars}{M.M.~Dworetsky, F.~Castelli \and R.~Faraggiana}
{ASP Conf. Series}{44}{232}
\article{Mathys, G.}{1995}{\aaa}{293}{746}
\article{Mathys, G., Hubrig, S.}{1995}{\aaa}{293}{810}
\article{Mathys, G., Hubrig, S.}{1997}{\aaas}{124}{475}
\article{Mathys, G., Lanz, T.}{1990}{\aaa}{276}{L21}
\inproceedings{Sargent, W.L.W., Searle, L.}
{1967}{Magnetic and Related Stars}{R.C. Cameron}{Mono Book Corporation}
{Baltimore}{209}
\article{Smith, K.C.}{1993}{\aaa}{276}{393}
\article{Smith, K.C.}{1994}{\aaa}{291}{521}
\article{Smith, K.C.}{1995}{\aaa}{305}{902}
\article{Smith, K.C.}{1997}{\aaa}{319}{928}
\article{Smith, K.C., Dworetsky, M.M.}{1993}{\aaa}{274}{335}
\article{Tokovinin, A.A.}{1993}{\sal}{19}{383}
\article{Tokovinin, A.A.}{1997}{\aaas}{124}{75}
\article{White, R.E., Vaughan, A.H., Preston, G.W., Swings, J.-P.}{1976}{\apj}
{204}{131}
\article{Zahn, J-P.}{1977}{\aaa}{57}{383}
\inproceedings{Zahn, J-P.}
{1984}{Observational Tests of Stellar Evolution Theory}{A.~Maeder \and
A.~Renzini}{IAU symp.}{105}{379}
\bibitem{}Zakharova, L.A.: 1997, Ph.D. Thesis 
%%%%%%%%%%%%%%%%%%%%%%%%%%%%%%%%%%%%%%%%%%%%%%%%%%%%%%%%%%%%%%%%%%%%%%%%%%%%%%
\end{thebibliography}
\end{document}